\documentstyle[12pt]{article}
\input html.sty
\input epsf
\begin{document}
\title{MATTERS OF GRAVITY, The newsletter of the APS Topical Group on 
Gravitation}
\begin{center}
{ \Large {\bf MATTERS OF GRAVITY}\\ 
\medskip
{\Large *******Anniversary Edition*******}}\\
\bigskip
\hrule
\medskip
{The newsletter of the Topical Group on Gravitation of the American Physical 
Society}\\
\medskip
{\bf Number 18 \hfill Fall 2001}
\end{center}
\begin{flushleft}

\tableofcontents
\vfill
\section*{\noindent  Editor\hfill}

\medskip
Jorge Pullin\\
\smallskip
Department of Physics and Astronomy\\
Louisiana State University\\
202 Nicholson Hall\\
Baton Rouge, LA 70803-4001\\
Phone/Fax: (225)578-0464\\
Internet: 
\htmladdnormallink{\protect {\tt{pullin@phys.lsu.edu}}}
{mailto:pullin@phys.lsu.edu}\\
WWW: \htmladdnormallink{\protect {\tt{http://www.phys.lsu.edu/faculty/pullin}}}
{http://www.phys.lsu.edu/faculty/pullin}\\
\hfill ISSN: 1527-3431
\begin{rawhtml}
<P>
<BR><HR><P>
\end{rawhtml}
\end{flushleft}
\pagebreak
\section*{Editorial}

Well, it is ten years since Peter Saulson put together the first
Matters of Gravity. A lot has happened since. The Topical Group was
formed and MOG became its official newsletter. As a celebration, we
decided to invite several prize-winners to give us a reflection of
their fields in the last and forthcoming decades. Finally, Gary
Horowitz, Carlo Rovelli (Xanthopoulos winners), Jens Gundlach (Pipkin
prizewinner), our former editor Peter Saulson and TGG founder Beverly
Berger accepted the challenge.

Otherwise not much to report here. If you are burning to have Matters of 
Gravity with you all the time, the newsletter is now available for
Palm Pilots, Palm PC's and web-enabled cell phones as an
Avantgo channel. Check out 
\htmladdnormallink{\protect {\tt{http://www.avantgo.com}}}
{http://www.avantgo.com} under technology$\rightarrow$science.
The next newsletter is due February 1st.
If everything goes well this newsletter should be available in the
gr-qc Los Alamos archives 
(\htmladdnormallink{{\tt http://xxx.lanl.gov}}{http://xxx.lanl.gov})
under number gr-qc/0109034. To retrieve it
send email to
\htmladdnormallink{gr-qc@xxx.lanl.gov}{mailto:gr-qc@xxx.lanl.gov}
with Subject: get 0109034
(numbers 2-17 are also available in gr-qc). All issues are available in the
WWW:\\\htmladdnormallink{\protect {\tt{http://www.phys.lsu.edu/mog}}}
{http://www.phys.lsu.edu/mog}\\ 
A hardcopy of the newsletter is
distributed free of charge to the members of the APS
Topical Group on Gravitation upon request (the default distribution form is
via the web) to the secretary of the Topical Group. 
It is considered a lack of etiquette to
ask me to mail you hard copies of the newsletter unless you have
exhausted all your resources to get your copy otherwise.
\par
If you have comments/questions/complaints about the newsletter email
me. Have fun.
\bigbreak

\hfill Jorge Pullin\vspace{-0.8cm}
\section*{Correspondents}
\begin{itemize}
\item John Friedman and Kip Thorne: Relativistic Astrophysics,
\item Raymond Laflamme: Quantum Cosmology and Related Topics
\item Gary Horowitz: Interface with Mathematical High Energy Physics and
String Theory
\item Richard Isaacson: News from NSF
\item Richard Matzner: Numerical Relativity
\item Abhay Ashtekar and Ted Newman: Mathematical Relativity
\item Bernie Schutz: News From Europe
\item Lee Smolin: Quantum Gravity
\item Cliff Will: Confrontation of Theory with Experiment
\item Peter Bender: Space Experiments
\item Riley Newman: Laboratory Experiments
\item Warren Johnson: Resonant Mass Gravitational Wave Detectors
\item Stan Whitcomb: LIGO Project
\end{itemize}
\vfill
\pagebreak

\section*{\centerline {
We hear that...}}
\addcontentsline{toc}{subsubsection}{\it  
We hear that... by Jorge Pullin}
\begin{center}
Jorge Pullin
\htmladdnormallink{pullin@phys.lsu.edu}
{mailto:pullin@phys.lsu.edu}
\end{center}

{\bf Robert Wald} has been elected to the National Academy of Sciences. 

{\bf Jens Gundlach} won the Pipkin award of APS ``For identifying, and
providing a solution to, an unrecognized weakness in the Cavendish technique
for measuring the gravitational constant G; improving the accuracy of G by 
an  order of magnitude, representing one of the largest incremental increases
in accuracy ever obtained in the history of such measurements.''

{\bf James Faller} won the Keithley award of APS ``For the development
of sensitive gravitational detectors and their successful application
to the study of physics and geophysics.''

{\bf Lawrence Krauss} won the Lilienfeld prize of APS ``For
outstanding contributions to the understanding of the early universe,
and extraordinary achievement in communicating the essence of physical
science to the general public.''

{\bf Juan Maldacena} won the Basilis Xanthopoulos award.

Hearty congratulations!

\vfill
\pagebreak
\section*{\centerline {
Matters of Gravity and the Topical Group in Gravitation}}
\addtocontents{toc}{\protect\medskip}
\addtocontents{toc}{\bf Reflections of a decade:}
\addtocontents{toc}{\protect\medskip}
\addcontentsline{toc}{subsubsection}{\it  
Matters of Gravity and the Topical Group in Gravitation, by Beverly Berger}
\begin{center}
Beverly Berger, Oakland University
\htmladdnormallink{berger@oakland.edu}
{mailto:berger@oakland.edu}
\end{center}

As ``Matters of Gravity'' celebrates its 10th year as the ``voice'' of
gravitational physics, it also celebrates its 7th year as the official
newsletter of the American Physical Society's Topical Group on
Gravitation (TGG).

The first mention of the TGG actually appeared in the Spring 1994 issue
of MOG with my open letter to the gravitational physics community stating
that  ``the time has come to create a TGG within the APS ... Such a group
would allow us to define and promote our interests and enhance our
visibility within the larger community of physicists. Construction of
LIGO and the Grand Challenge Supercomputing Project on the two black hole
problem are only two examples of the significant developments in our
field that render such advocacy essential.'' The required 200 signatures
to petition the APS to form the TGG were eventually obtained. The TGG came
into existence when its formation was approved by the APS Council in
April 1995.

All units (divisions, topical groups, forums) within the APS are supposed
to have newsletters. It was natural, therefore, that the
well-established, well-regarded, and reliably published MOG become the
newsletter of the TGG. In fact, the one exception made to the ``plain
vanilla'' bylaws of the TGG was to allow the editor of the newsletter to
be someone other than the Secretary / Treasurer of the topical group.
Thus Jorge Pullin was prevailed upon to continue his excellent
stewardship of the newsletter. The first ``official'' TGG issue of MOG
appeared in Fall 1995.

Since its beginnings in 1995, the TGG has been an unqualified success.
Its membership has grown from its initial 200 or so to approximately 550.
The TGG has increased the visibility of gravitational physics within
physics by its active participation in the APS April Meetings, starting
with the 1996 meeting. This has been achieved by the TGG's sponsorship of
sessions of invited talks on recent developments in gravitational physics
both on its own and jointly with the Divisions of Astrophysics, Particles
and Fields, Computational Physics, and the Topical Group in Fundamental
Constants and Precision Measurement. For the first time this past April,
the TGG provided travel grants to encourage students to attend the April
Meeting.

Many distinguished gravitational physicists have been elected to the TGG
leadership. Past chairs include Kip Thorne, Abhay Ashtekar, Rai Weiss, and
Cliff Will. Jim Isenberg was the first Secretary / Treasurer with Jim
Bardeen, Sam Finn, Leonard Parker, Fred Raab, David Shoemaker, Bob Wald,
Mac Keiser,Steve Carlip, Peter Saulson, John Friedman, and
Bob Wald serving on the Executive Committee. The current officers are
Bob Wald (Chair), Richard Price (Chair Elect), John Friedman (Vice
Chair), and David Garfinkle (Secretary / Treasurer). The current
executive committee consists of Ted Jacobson, Jennie Traschen, Eanna
Flanagan, Gabriela Gonzalez, and Matt Choptuik.

In what we hope will be the very near future, the APS will award the
first Einstein Prize in Gravitational Physics. This prize was developed
due to the initiative of the TGG leadership. All APS prizes must be
approved by the APS Council (including the name of the prize) and must
have a substantial endowment. After some discussion over the naming of
the prize, the APS Council authorized the TGG to conduct a fundraising
campaign for the prize's endowment. This campaign is in progress.

The TGG has also exercised its right, as an APS unit, to name APS
Fellows. Since the TGG formed, it has named approximately 10 fellows.

More information about the TGG may be found at its web page, and in
the various issues of MOG.

\htmladdnormallink{\protect{http://gravity.phys.psu.edu/\~{}tggweb/}}
{http://gravity.phys.psu.edu/\~{}tggweb/}

\vfill
\pagebreak
\section*{\centerline {
10 Years in Gravitational Wave Detection}}
\addcontentsline{toc}{subsubsection}{\it  
10 Years in Gravitational Wave Detection, by Peter Saulson}
\begin{center}
Peter Saulson, Syracuse University
\htmladdnormallink{saulson@phy.syr.edu}
{mailto:saulson@phy.syr.edu}
\end{center}

Looking back ten years in the field of gravitational wave detection is
especially handy, since the life span of Matters of Gravity roughly
coincides with the life span of LIGO (the Laser Interferometer
Gravitational Wave Observatory, in case anyone doesn't know) as an
approved project. There is of course a long pre-history to LIGO -- the
inception of the idea, planning and feasibility studies, technology
development, and, not least, lobbying the physics community, the NSF,
and Congress to earn approval. Still, life changed for all of us when
LIGO was approved.

I vividly remember the moment when I learned that the construction of
LIGO had been approved. Abhay Ashtekar, who has always been much
better plugged in than I, greeted me in the hallway of the Physics
Building at Syracuse University with an outstretched hand and a big
grin. As he shared the news, I confess that my own feeling was not the
unalloyed happiness that showed on his face. I think I mumbled
something like, "Oh, shit, now we have to make it work." It was a time
in my life when the technological hurdles facing LIGO seemed
especially daunting. Today, when a great many bullets have already
been dodged, the fear that I felt seems much less justified.

While I'm confessing my old fears, perhaps I should mention another
moment when the enormity (literally) of what we have taken on in LIGO
came home to me. It was several years later, during my first visit to
the LIGO site at Hanford WA, when I walked through the then-empty
experimental hall at the vertex, euphoniously called the LVEA (Laser
and Vacuum Equipment Area.) Even more than the 4 km arms, the vastness
of this workspace brought home to me the magnitude of what we had
persuaded the American taxpayer to support. Now that the Hanford LVEA
and the one at its sister site at Livingston LA have been filled with
their vacuum chambers and then in turn with the optics for LIGO's
interferometers, the scale seems fully appropriate. LIGO is big, but
it is big for a good reason - to maximize our chances of detecting
gravity waves.

Of course, the detection of gravity waves remains to be
accomplished. But much along the way has been done successfully, and
with a certain amount of style. The aforementioned remote laboratories
have been built, equipped, and staffed. All three of LIGO's
interferometers have now been installed. They are in various stages of
commissioning; the 2 km interferometer at Hanford (the first of the
three to be installed) has "locked" in essentially its full servo
configuration, and the 4 km interferometer at Livingston has gone
almost as far. The servo engineering that has made this possible is a
real tour de force, with gains (and signs!) of feedback switching as
the interferometer progresses through a series of states approaching
the full Power-Recycled Fabry-Perot Michelson configuration. A few
years ago we didn't know how to do this, but now it works.

Now that the interferometers are moving into a state where they
function, work is commencing on understanding their performance. It
must be said that at present the noise levels are substantially poorer
than the design performance of the initial LIGO interferometers. But
some of the reasons are well understood, so it is reasonable to hope
that what is a huge gap at the present will start to close rapidly. Of
course, it is hard to predict how quickly the last order of magnitude
will be crossed.

This technical progress could only have come about through progress in
the social organization of LIGO. The first proposal for LIGO in 1987
(the one before the successful 1989 proposal) listed 18 members of the
team. Now, a scan of the LIGO roster reveals over 180 names of staff
at Caltech, MIT, Hanford and Livingston (including only one name of
the original 18.) Growth of this magnitude could not have been
effectively managed without leadership by people experienced with
large projects. This expertise came to LIGO in the person of Barry
Barish and of the colleagues he brought with him to LIGO from high
energy physics in the middle of the last decade.

The social/scientific structure has grown in another important way as
well. LIGO is now organized into two bodies. The staff above
constitute the LIGO Laboratory, the group responsible for ensuring
that the LIGO interferometers function properly. Direction of LIGO's
scientific program lies in a larger body called the LIGO Scientific
Collaboration (or LSC), consisting (in a recent count) of 112
scientists, engineers, and technicians from within the Lab, plus an
additional 239 members from 27 other groups from across the U.S. and,
indeed, the rest of the world. And the LSC continues to grow; at its
last meeting in August, two new groups joined.

In addition to the dramatic growth of the number of people working on
LIGO, another important change has taken place in the style of
work. With the installation of the LIGO interferometers at the
Observatories, the focus of work has shifted toward the sites. In
addition to the staff that have moved to Washington or Louisiana, this
has meant a great deal of travel by experts based at Caltech and
MIT. Some non-Lab LSC groups have also been a big presence at the
sites. I had a chance to view this process first-hand in 2000 when I
spent a sabbatical year at the Livingston Observatory. A substantial
chunk of the important work was being done by the group from the
University of Florida, who supplied the interferometers' Input
Optics. Another important presence was that of the growing group at
neighboring LSU, who are vital participants in the commissioning work.

Looking back, I must confess that one of my most vivid impressions
from that year at Livingston was yet another epiphany of the magnitude
of LIGO's work. Early on, I realized that we needed a set of design
documents for reference while we worked on commissioning the
interferometer. I searched LIGO's on-line Document Control Center, and
by the time I was done I had found a bookshelf full of subsystem
descriptions, analyses, and plans. This brought home to me not only
the complexity of LIGO as a scientific instrument, but the remarkable
intensity and quality of the work it has taken to produce it. Visit a
LIGO site and marvel at its size, as I did at first; then pause to
admire even more the richness of labor it has taken to turn that site
into a gravitational wave detector.

Even as commissioning of the interferometers goes on, work has been
progressing on preparing to collect and analyze data to search for
gravitational waves. Here is another massive effort, largely invisible
to outsiders, that has almost completed the data analysis system that
will enable the 24/7 search for the gravitational wave needles in the
haystacks of data that LIGO will produce.

Over the past year or two, the interferometers have been exercised in
a set of Engineering Runs that practiced collecting data for extended
periods, while the software has been tested in a set of Mock Data
Challenges that practiced analyzing the data. These parallel efforts
will come together in an Upper Limit Run, now scheduled for around New
Years. We will run the interferometers for two weeks, collect the
data, and analyze that data to the point of being able to make
scientific claims about the presence (or, more likely, absence) of
gravitational wave signals in it. Throughout 2002, LIGO plans to
intersperse interferometer improvement with data-taking periods, until
full-time operation at design sensitivity is achieved.

While all this has been going on, LIGO has been preparing increasingly
detailed plans for a new set of interferometers to be installed at the
sites after the initial LIGO Science Run has been completed. Advanced
LIGO will have roughly an order of magnitude better strain sensitivity
than the initial interferometers, which ought to be enough to
guarantee that known sources (especially binary neutron star
inspirals) will be within reach of detection.

Most of this review has focused in a rather parochial fashion on the
American effort, but it needs to be stressed that equally impressive
work is going on at a number of other places around the world. A
Japanese 300-meter interferometer, TAMA, has already started
operating, and is showing remarkable performance. The British-German
GEO 600-meter interferometer near Hannover is almost completely
installed; it features advanced mirror suspensions and optics that
will pave the way for parts of Advanced LIGO. The 3-km VIRGO
interferometer near Pisa (a joint French-Italian project) is also well
along; it is demonstrating advanced seismic isolation systems that
will let it probe to lower frequencies than any other instrument yet
built. One other admirable development should be stressed - the
remarkable extent to which scientific cooperation is being maintained
between all of these nominally competing efforts. LIGO and GEO have
become especially close, with the GEO team having joined the LSC and a
full reciprocal data exchange agreement having recently been
concluded.

The past decade has seen another remarkable occurrence on the way to
the establishment of the science of gravitational wave detection. An
interferometer in space has been a gleam in a few pairs of eyes for
almost as long as LIGO has, but now it appears well on the road to
becoming a reality. The LISA project garnered strong European support
several years ago, and appears to be marching inexorably toward
becoming a key part of the NASA program as well. With a projected
sensitivity that easily places it in a position to study a range of
interesting sources, LISA has also been blessed with a remarkable
degree of support from the astronomical community.

With any luck, the next 10 year review will be able to look back on
the detection and study of gravitational waves.

\vfill\eject
\section*{\centerline {Ten years of general relativity, some reflections}}
\addcontentsline{toc}{subsubsection}{\it  
Ten years of general relativity, some reflections, by Carlo Rovelli}
\begin{center}
Carlo Rovelli, Centre de Physique Theorique, Marseille
\htmladdnormallink{carlo@rovelli.org}
{mailto:carlo@rovelli.org}
\end{center}

Ten years of {\em Matters of Gravity\/} (thanks a lot, Jorge!), ten
years of research in general relativity.  It has been a period of
triumph for GR: We have seen evidence for gravitational emission from
binary pulsars, with theory and observations matching with a level of
accuracy previously found only in quantum field theory.  We have been
enchanted by the gravitational lensing images.  Black holes have moved
from exotic theoretical hypothesis to realistic objects in the sky. 
Last month, I got lost in a trip, and could trace my way back thanks
to a small electronic device that wouldn't work without taking GR's
corrections into account.  Particle physicists, that not long ago kind
of looked down at GR, nowadays use Einstein's space-times in every
other paper.  Observational cosmology has exploded, relativistic
astrophysics is solidly established; the problem of quantizing the
theory is now recognized as perhaps the main problem in theoretical
physics, and the papers devoted to it keep growing.  Large resources
are being invested internationally in the search for gravitational
waves and computational GR \ldots\ \ The list of successes could
continue.  How long a road from the sleepy and a bit esoteric GR
community, when black holes were badly understood exotic theoretical
hypotheses, the only experimental support came from the three
``classic tests" and the rest of physics looked at us with suspicion,
if it payed any attention at all.  It has been a breathtaking decade.

Many problems remain open, and so much remains to be done.  And there
are also some dangers ahead.  We are all holding our breath waiting
for the gravitational waves.  We are solidly beyond our colleagues
involved in this adventure and we are optimistic, but also a bit
concerned: the community has taken some risks with this search, and if
it took too long, it wont be good for all of us.  Great
efforts are been put in computational GR. Again, let's hope for the
better, but we should keep in mind the experience with computational
lattice QCD, where great skills and money were expended, and great
hopes raised, with far less results than hoped for.  Excitement is high
in my own field, loop quantum gravity, where the feeling is that
perhaps we are having true glimpses into the quantum structure of
spacetime.  But let us not forget how many tentative quantum theories
of gravity have claimed victory and then were proven unsatisfactory.

The worst danger, I think, is that theoretical physics, and sometimes
even experimental physics, is nowadays often so far removed from the
actual final experimental outcome (the only final arbiter), that the
temptation is dangerously high to keep selling for good whatever we
have.  I am afraid that some portion of physics have moved down this
dangerous path.  But success, I think, can only be granted by
scrupulous intellectual honesty.  The high respect and the credibility
that science enjoys rely on the intellectual honesty of the
scientists.  Several people now begin to suspect that something has
got wrong on this in the last decade.  At a recent conference I had a
conversation with a brilliant young researcher.  In the conversation,
two theories were mentioned: GR and some particular supersymmetric
theory in high dimension.  Casually, I said that at least we knew that
one of the two was experimentally supported.  My young friend asked
which one.  I thought he was joking, but he was not.  In his mind
there was absolutely no understanding of the distinction between a
theory whose novel peculiar predictions have found a huge wealth of
empirical support, and a complex theoretical hypothesis that for the
moment has no empirical support whatever.

The distinction between what we have learned about the world on the
one hand, and our attempts to understand more on the other hand, is
the rock over which science bases its strength.  I am afraid that this
distinction is becoming a bit obfuscated in some areas of theoretical
physics, and hypotheses are too often sold for facts.  This may
increase funding, positions and political power in the short run, but
it is a recipe for disaster in the long run.  I think that the
theoretical physics community should seriously react against this
attitude, which is endangering its own position in the world.  I have
heard many scientists repeating this privately in the corridors of the
conferences.  Perhaps they should say it more vocally and more
publicly.

\vskip.5cm

As a result of the successes of GR (and also of the overwhelming and
unexpected empirical success of the particle physics standard model),
the relation between the GR community and the rest of physics has much
changed in this decade.  Ten years ago, the divide between the GR
community and the rest of theoretical physics was sharp.  Outside our
small community, spacetime was unquestionably flat and non dynamical. 
Today, some basic ideas of GR pervade large parts of theoretical
research.  GR is being finally universally accepted as a component of
our present understanding of the world.

But, as we know well, GR is much more than a theory of gravity, namely
much more than the specific theory for a specific physical
interaction.  It is a rethinking of the notions of space and time,
which involves the entirety of our understanding of the world.  In my
opinion, the deepness and the richness of the shift in perspective
produced by GR is far from being fully understood and fully absorbed. 
GR does not claim only that spacetime is curved and satisfies certain
equations.  Rather the most far reaching physical consequence of the
theory, which follows from the invariance properties of its equations,
is the discovery that no physical meaning can be attached to the
coordinates, and physical localization can therefore only be defined
relationally.  Dynamical objects are physically localized only with
respect to each others.  This is a huge conceptual jump out of
Newtonianism, which brings our understanding of spacetime back to
Cartesian (and Aristotelian) relational notions of space.  The
Newtonian localization with respect to space (that allows Newton to
define acceleration as absolute) is reinterpreted in GR as
localization with respect to a particular dynamical object: the
gravitational field.

In my opinion, Einstein's discovery that the gravitational field and
the spacetime metric are the same entity, is not well expressed by
saying that there is no gravitational field, just a curved spacetime. 
Rather, it is better expressed by saying that there is no spacetime,
just the gravitational field.  The gravitational field is,
dynamically, a field like the others.  But the fields do not live over
a spacetime, they leave, so to say, over each other.

I think that this profound change of perspective on the world, has not
yet been completely absorbed.  The hardest part to digest is not the
relational nature of space; it is the relational nature of time.  To
this, many instinctively resist.  Giving up the idea of an external
flowing time along which things happen is hard, as it was hard giving
up the idea of the center of the universe, or the idea of absolute
rest.  I am convinced that this change of perspective reflects a
deeper understanding of the physical structure of the world and will
stay with us for a while in the physics of the future.

But while the Einstein equations are being widely used in fundamental
physics, this conceptual revolution is still little understood. 
World famous theoreticians still search the fundamental theory over a
background Minkowski spacetime (perhaps in high dimensions).  In my
opinion, they have not understood what we have learned about the world
with GR.

Large sectors of basic physics still expect to be thought again from
scratch at the light of this conceptual revolution.  Classical
Hamiltonian mechanics has proven flexible enough to consistently
extend to general covariant physics (where there is no canonical time
and no Hamiltonian).  But thermodynamics and statistical mechanics
still wait to find a formulation sufficiently general to take the GR
revolution into account.  And of course, so does quantum theory.  In
the XXth century, quantum theory and general relativity have changed
in depth our understanding of the world; we are still far from a
consistent picture of the physical world that can take the two
conceptual novelties into account.  The great scientific revolution
opened by the XXth century is not over: the cards are one the table
and expect to be put in the right order.  Could there be a more
exciting period for researching in fundamental physics?

\vfill\eject
\section*{\centerline{Tabletop gravity experiments}}
\addcontentsline{toc}{subsubsection}{\it  
Tabletop Gravity Experiments, by Jens Gundlach}
\begin{center}
Jens Gundlach, University of Washington
\htmladdnormallink{gundlach@npl.washington.edu}
{mailto:gundlach@npl.washington.edu}
\end{center}

The past fifteen years of laboratory-scale gravitational
experimentation have been marked by many new and exciting
developments. The field received a lot of impetus by the hypothesis of
a "fifth force" [1] in 1986. This very testable new force
would have been a blatant violation of the equivalence principle. The
evidence for the 5$^{th}$ force was partially based on a reanalysis of
the torsion balance data of Baron E\"otv\"os of the early
1900's. Immediately several groups around the world started to do
E\"otv\"os-type experiments. The availability of new technologies
combined with many new and creative ideas quickly led to several
refined measurements by which the 5$^{th}$ force in its postulated
form could be conclusively ruled out. However, the physics community
was once again reminded of the importance of the equivalence principle
which lies at the foundation of general relativity. Tests of the
equivalence principle become particularly important for grand
unification schemes, most of which predict an equivalence principle
breakdown at some level. In addition it is generally believed that the
standard model of particle physics can only be complete with the
existence of new particles which could exist at high masses as well as
at the ultra low energies. The latter frontier being covered by
laboratory-gravity tests.

Most equivalence principle tests compare the acceleration of different
materials towards another source mass. The difference in test mass
composition is chosen to maximize the new interaction's charge, which
could be e.g. baryon number, lepton number or combinations
thereof. The source mass could be a mass in the lab, a nearby hill,
mountains, the entire earth, the Sun, the Milky Way or even
cosmological structures. Several types of instruments were
developed. One of the more exotic devices consisted of a perfectly
buoyant hollow copper sphere in water tank placed at a
cliff [2]. Others compared the rate of free fall of
different masses [3]. By far the most sensitive and versatile
devices proved to be torsion balances. Here new concepts as well as
quantitative understanding led to tremendous advances. Our group at
the University of Washington, called the E\"ot-Wash group, developed a
torsion balance that is installed on a continously rotating
turntable. As seen from a restframe, turning with the turntable, the
signal is modulated at the rotation frequency of the turntable. The
technical difficulty lay in producing the required extremely constant
rotation rate. We also introduced a multipole analysis that proved
very practical in eliminating gravitational torques that could have
been mistaken for an equivalence principle violation. The differential
acceleration sensitivity between different materials that we are now
achieving is $\Delta a < 5\times10^{-15}m/s^{2}$. This limits
equivalence principle violations with infinite range and baryon number
as its charge to be at least $10^{9}$ times weaker than
gravity. Together with another experiment, in which a 3 ton source was
rotated about a stationary pendulum, we now can set new limits on
equivalence principle violations for ranges from the
cm-range [4] to infinity [5]. Riley Newman's group at UC
Irvine also has a long and successful tradition of torsion balance
experiments probing gravity. He has pioneered cryogenic torsion
balances that will have phenomenal intrinsic sensitivity
[6].

In the last few years the $1/r^2$-law of gravity at very short ranges
came under close scrutiny. Several theorists [7] argued that
it might be possible for some of the unobserved extra dimensions in
string theory to be compactified close to a mm-radius rather than at
Planck length. For two such dimensions the $1/r^2$-force law would
break down below the mm-scale, precisely at a length range where
limits from previous experiments were weak. A group at the University
of Colorado and another group at Stanford University built
micromechanical oscillator plates which would be brought into
resonance by a close-by parallel moving source plate if the
$1/r^2$-law were violated. Both groups use sophisticated mechanical
vibration isolation techniques, as well as an electrostatic shield
between the source and the sensor. Our approach involved a torsion
balance. We built a pendulum consisting of a horizontal disk with 10
holes drilled in it. Below the pendulum we located a similar
horizontal disk also with 10 holes. This source disk was mounted on a
slowly rotating turntable. Gravity causes the pendulum to be deflected
10 times per revolution. We placed another disk below the source disk
that has 10 holes exactly out of phase with the upper disk. This disk
was designed to exactly cancel the gravity signal, assuming $1/r^2$
holds. With this setup we were able to tell that a $1/r^2$-violation
must have a Yukawa range shorter than $\approx$0.2mm for a strength
about equal to gravity [8].

Contrary to the equivalence principle and the $1/r^{2}$-tests several
new measurements of the gravitational constant G were motivated by a
disagreement in experimental results. One well respected measurement
deviated by $\approx42\sigma$ from the accepted value. This situation
forced an increase in the uncertainty of the accepted value of G by a
factor of 12 (now 0.15\%) [9]. In addition Kuroda [10]
discovered that torsion fiber anelasticity, a material property, had
led to a bias in many previous measurements.  Several measurements
were initiated, each with new approaches to minimize systematic
uncertainties. Torsion balances continued to dominate. Using our
experience from the equivalence principle tests we built a continuously
rotating balance. Uncertainties with the torsion fiber were avoided by
regulating the turntable velocity so that the fiber was not
twisted. The gravitational signal was derived from the turntable
acceleration. We discovered that a thin vertical plate pendulum
eliminated the difficult pendulum metrology issues most measurements
had. Rotating the attractor masses on a coaxial turntable transformed
our signal to a higher frequency. Our result is about 250ppm higher
than the accepted value and has an uncertainty of
14ppm [11].  Another group [12]  led by Terry Quinn at
the BIPM in Paris eliminates the anelasticity problem by using a
torsion strip instead of a round fiber. A four-fold attractor-pendulum
configuration is used. The likelihood of unknown systematic error is
reduced by using two independent torque measurements: electrostatic
feedback and a calibrated deflection. Their result has been submitted
for publication.  Riley Newman's group has operated a torsion balance
at 2K [13]. The group was able to show that at these
temperatures anelasticity corrections are small and well
understood. They also use a flat plate pendulum. Two copper rings as
attractors simplify their metrology issues. The apparatus is located
at a remote site to reduce noise. The group expects to announce
results soon.

{\bf References:}

\noindent[1] E. Fishbach et al., Phys. Rev. Lett. {\bf 56}, 3 (1986).  

\noindent[2] P. Thieberger, Phys. Rev. Lett. {\bf 58}, 1066 (1987).  

\noindent[3] K. Kuroda and N. Mio, Phys. Rev. {\bf D42}, 3903 (1990),
T.M. Niebauer, M.P. McHugh, J.E. Faller, Phys. Rev. Lett. {\bf 59},
609 (1987).

\noindent[4] G. L. Smith et al., Phys. Rev. {\bf D61}, 022001 (1999).

\noindent[5] Y. Su et al., Phys. Rev. {\bf D50}, 3614 (1994).

\noindent[6] M.K. Bantel and R.D. Newman, Class. Quantum Gravity {\bf
17}, 2313 (2000).

\noindent[7] For example: N. Arkani-Hamed, S. Dimopoulos, and
G. Dvali, Phys. Lett {\bf B429}, 263 (1998).

\noindent[8] C.D. Hoyle et al., Phys. Rev. Lett. {\bf 75}, 2796 (2001).

\noindent[9] P.J. Mohr and B.N. Taylor, J. Phys. Chem. Ref. Data {\bf
28}, 1713(1999).

\noindent[10] K. Kuroda, Phys. Rev. Lett. {\bf 75}, 2796 (1995).

\noindent[11] J.H. Gundlach and S.M.~Merkowitz, Phys. Rev. Lett. {\bf
85}, 2869 (2000).

\noindent[12]  T. Quinn et al., Meas. Sci. Technol. {\bf 10}, 460 (1999).

\noindent[13] R. Newman and M. Bantel, Meas. Sci. Technol. {\bf 10},
445 (1999).

\vfill\eject
\section*{\centerline{String Theory: The Past Ten Years}}
\addcontentsline{toc}{subsubsection}{\it  
String Theory: The Past Ten Years, by Gary Horowitz}
\begin{center}
Gary Horowitz, UC Santa Barbara
\htmladdnormallink{gary@cosmic.physics.ucsb.edu}
{mailto:gary@cosmic.physics.ucsb.edu}
\end{center}

Given a choice between summarizing the past decade of achievements in string
theory or speculating about what string theory might look like a decade from
now, I have chosen the first option. Indeed,  given the rapid progress over
the last decade, I find it hard to guess where
string theory will be even a few years from now.

Ten years ago, it was common (and correct) to distinguish the two main
approaches to quantum gravity by saying that string theory [1]
was   perturbative,
and background dependent while the other approach [2] was
non-perturbative and
background independent. In light of this, it is not surprising that most
relativists were not interested in string theory.
Today, this distinction is no longer applicable.
As we will discuss, there is now a complete, non-perturbative and
background independent formulation of the theory, at least for space-times
with certain asymptotic boundary conditions.

Let me begin by summarizing the situation ten years ago. At that time, there
were five perturbatively consistent string theories. They
were all based on the idea that particles are just different excitations of
a one-dimensional extended object -- the string. They all included gravity,
supersymmetry, and required ten spacetime dimensions. These theories
differed in the
amount of supersymmetry and type of gauge groups that were included.
In addition to perturbations about Minkowski spacetime, nontrivial classical
solutions were known, including space-times in which six of the spatial
dimensional are compactified. In some cases, the resulting four dimensional
effective theories
were in qualitative agreement with observations.
It was also known that spacetime is seen
differently in string theory than in general relativity or
ordinary field theory. In particular, flat spacetime with one direction
compactified into a circle of radius $R$ is completely equivalent to a
spacetime with a circle of radius $\ell_s^2/R$ where $\ell_s$ is a new
dimensional parameter related to the string tension. This is possible since
the string is an extended object and has winding states in addition to the
usual momentum states.

One of the main things that has changed over the past decade is that we now
know that string theory does not just involve strings. Higher (and lower)
dimensional objects (called branes) play an equally fundamental role.
Using these branes, convincing evidence has been accumulated that all five
of the perturbative string theories are just different limits of the same
theory, called M theory. (There is no agreement about what the M stands
for.)
There is yet another limit in which M theory reduces
to {\it eleven dimensional} supergravity.

Without a doubt, the main achievement of string theory over the past decade
has been an explanation of black hole entropy [3]
For a class of near extremal
four and five dimensional charged black holes (with the extra spatial
dimensions
compactified on e.g. a torus) one can count the number of microstates of
string theory associated with the black hole. One finds that in the limit
of large black holes, the number is exactly the exponential of the
Bekenstein-Hawking entropy. The black holes can include angular momentum,
and
several different types of charges, so the entropy is a function of
several parameters. The string calculation reproduces this function exactly.
Even more surprising, it was shown that the radiation calculated in string
theory agrees exactly with the Hawking radiation from the black hole,
including the distortions of the thermal spectrum arising from the greybody
factors [4].

By exploring these black hole results, Maldacena was led to his famous
``AdS/CFT" conjecture [5].
This states that string theory (or M theory) on
space-times which asymptotically approach anti de Sitter (AdS) space
is completely described by a conformally invariant
field theory (CFT) which lives
on the boundary of this spacetime. This is a remarkable conjecture which
states that an ordinary field theory in a fixed spacetime
can describe all of string theory with asymptotically
AdS boundary conditions. Since only the asymptotic
boundary conditions on the metric
are fixed, this constitutes a background independent formulation of the
theory. Since the CFT can be defined non-perturbatively, this is also a
non-perturbative formulation.
The AdS/CFT conjecture
is a concrete implementation of the idea that quantum gravity
should be ``holographic" [6], i.e., the true degrees of freedom live
on the boundary, but can describe all physical processes in the bulk.
This was originally suggested by the fact that
black hole entropy is given by the horizon area, but now  applies to all
quantum gravity processes, not just black holes. This conjecture has
withstood a number of nontrivial checks.
It can be
used to derive new predictions about
strongly coupled CFT, or learn about quantum gravity. For example,
one immediate consequence is that the formation and evaporation of a small
black hole in AdS can be described by the unitary evolution of a state in
the CFT.

There is much that remains to be done.
Major open questions include: (1) Develop a dictionary to translate
spacetime
concepts into field theory language and vice versa. (Only a few entries
in this dictionary are currently known.) In
particular, find a ``spacetime reconstruction theorem" which allows us
to reconstruct a semiclassical spacetime from certain states in the CFT.
(2) Extend the AdS/CFT conjecture to other
boundary conditions including asymptotically flat space-times. (3) Calculate
the entropy of all black holes (including Schwarzschild) exactly and
understanding why it is always proportional to the horizon area.

{\bf References:}

[1]  J. Polchinski, {\it String Theory}, in 2 vols., Cambridge
Univ.
Press (1998).

[2]  C. Rovelli, ``Loop Quantum Gravity", Living Reviews 1 (1998),
\htmladdnormallink{gr-qc/9710008}{http://xxx.lanl.gov/abs/gr-qc/9710008}

[3]  A. Strominger and C. Vafa, ``Microscopic Origin of the
Bekenstein-Hawking Entropy", Phys. Lett. B379 (1996) 99,
\htmladdnormallink{hep-th/9601029}{http://xxx.lanl.gov/abs/hep-th/9601029}; 
G. Horowitz, ``Quantum States of Black Holes", in {\sl
Black Holes and Relativistic Stars}, ed. R. Wald, U. of Chicago Press
(1998),
\htmladdnormallink{gr-qc/9704072}{http://xxx.lanl.gov/abs/gr-qc/9704072};
 A. Peet, ``TASI Lectures on Black Holes in String Theory",
\htmladdnormallink{hep-th/0008241}{http://xxx.lanl.gov/abs/hep-th/0008241}.

[4] J. Maldacena and A. Strominger, ``Black Hole Greybody
Factors and D-Brane Spectroscopy", Phys. Rev. D55 (1997) 861.

[5] J. Maldacena, ``The Large N Limit of Superconformal Field Theories
and Supergravity", Adv. Theor. Phys.  2 (1998) 231,
\htmladdnormallink{hep-th/9711200}{http://xxx.lanl.gov/abs/hep-th/9711200};
O. Aharony, S. Gubser, J. Maldacena, H. Ooguri, and Y. Oz, ``Large N
Field Theories, String Theory and Gravity", Phys. Rept.  323 (2000)
183,
\htmladdnormallink{hep-th/9905111}{http://xxx.lanl.gov/abs/hep-th/9905111}.

[6] G. 't Hooft, ``Dimensional Reduction in Quantum Gravity",
\htmladdnormallink{gr-qc/9310026}{http://xxx.lanl.gov/abs/gr-qc/9310026};
L. Susskind, ``The World as a Hologram", J. Math.  Phys. 36 (1995)
6377,
\htmladdnormallink{hep-th/9409089}{http://xxx.lanl.gov/abs/hep-th/9409089}.

\vfill\eject

\section*{\centerline {
Fourth Capra Meeting on Radiation Reaction}}
\addtocontents{toc}{\protect\medskip}
\addtocontents{toc}{\bf Conference Reports}
\addtocontents{toc}{\protect\medskip}
\addcontentsline{toc}{subsubsection}{\it  
Fourth Capra Meeting on Radiation Reaction, by Lior Burko}
\begin{center}
Lior Burko, California Institute of Technology
\htmladdnormallink{burko@tapir.caltech.edu}
{mailto:burko@tapir.caltech.edu}
\end{center}

The Capra meetings on radiation reaction are annual gatherings, the
fourth of which was hosted by Carlos Lousto at the
Albert-Einstein-Institut in Golm, Germany, May 28--31, 2001.  The
first meeting in this series was held in 1998 at a ranch in
northeastern San Diego county in California. This ranch was once owned
by Frank Capra, the director of such movies as ``Mr.\ Smith Goes to
Washington'' and ``It's a Wonderful Life''. Capra was a Caltech
alumnus, and donated the ranch to Caltech. In the tradition of the
``Texas'' meetings, each meeting in this series is called a ``Capra''
meeting, even if the venue is rather removed from Capra's
ranch. Summaries of previous Capra meetings appeared in Matters of
Gravity, No.\ 14 (Fall 1999) (Capra2 by P.\ Brady and A.\ Wiseman) and
No.\ 16 (Fall 2000) (Capra3 by E.\ Poisson).

The Capra meetings focus on radiation reaction and self interaction in
general relativity. The motivation for this topic is twofold. First,
the two-body problem in general relativity is as yet an unresolved
problem.  Even the restricted two-body problem, where the mass ratio
of the two bodies is extreme, lacks in understanding. It is this
restricted problem which the Capra meetings focus on. In the limit of
infinite mass ratio, a test mass moves along a geodesic of the
spacetime created by the massive body. When the mass ratio is finite,
the energy-momentum of the small mass acts as an additional source for
spacetime curvature, which affects the motion of the small mass
itself. Specifically, the small mass now moves along a geodesic of a
perturbed spacetime. An alternative viewpoint is to construe the
motion of the small mass as an accelerated, non-geodesic motion in the
unperturbed spacetime of the big mass. This acceleration is then
caused by the self force of the small mass.  Although for many
interesting cases it is sufficient to restrict the discussion to
linearized perturbations (thanks to the high mass ratio), there still
remains an inherent difficulty: the metric perturbations typically
diverge at the coincidence limit of the field's evaluation point and
the source of the perturbations. It is the removal of this divergence,
or the regularization problem of the self force, which constitutes the
greatest hurdle in the solution of the restricted two-body problem.

The second motivation stems from the prospects of detecting
low-frequency gravitational waves with the Laser Interferometer Space
Antenna (LISA), which is currently scheduled to fly as early as
2010. One of the most interesting potential sources for LISA is the
gravitational radiation emitted by a compact object spiraling into a
super-massive black hole, like those in galaxy centers. The typical
mass ratio is then $10^{5-7}$, which makes the restricted two-body
problem relevant. During the last year of inspiral (the LISA
integration time) the system can undergo $(1-5)\times 10^5$ orbits. In
order to generate accurate templates which track the system over so
many orbits, it is required to compute the orbital evolution (due to
both dissipative and conservative effects) to high accuracy, which
requires the inclusion of self interaction.

Twenty talks, covering many aspects and approaches to the problem,
were given at the fourth Capra meeting. The following short
description is greatly biased by my own understanding and taste. A
full list of the talks, including the online proceedings of the
meeting (namely, links to the slides used by the speakers), appears at
the meetings web page:

\htmladdnormallink{\protect {http://www.aei-potsdam.mpg.de/\~{}lousto/CAPRA/Capra4.html}}
{http://www.aei-potsdam.mpg.de/\~{}lousto/CAPRA/Capra4.html}

A number of different approaches for the calculation of the self force
have been suggested. These are approaches for the computation of the
``tail'' part of the self force [1,2]. M.\ Sasaki, Y.\
Mino, and H.\ Nakano presented progress obtained in Power-Expansion
Regularization. M.\ Sasaki described, in addition to Power-Expansion
Regularization also an alternative approach of Mode-by-Mode
Regularization, and also discussed the problems of extending the work
to Kerr background, and the difficult gauge problem. Y.\ Mino
described in great detail the mathematical techniques which are needed
for Power-Expansion Regularization. H.\ Nakano showed how to apply
this approach for the computation of the self force acting on a scalar
charge in circular orbit around a Schwarzschild black hole
[3]. L.\ Barack described his work with A.\ Ori on the
extension of Mode-Sum Regularization to the gravitational case
[4], and L.\ Burko discussed work with Y.-T.\ Liu on how
Mode-Sum Regularization can be applied for the case of a static scalar
charge in the spacetime of a Kerr black hole, even without knowledge
of the Mode-Sum regularization function [5]. W.\
Anderson presented progress obtained with \'{E}.\ Flanagan and A.\
Ottewill in the approach of normal neighborhood expansion.  A.\ Ori
presented work with E.\ Rosenthal on extended-body models, and showed
how to re-derive the Abraham-Lorentz-Dirac equation (in flat
spacetime) using such models, based on momentum considerations. This
approach appears to be very promising also in curved spacetime. C.\
Lousto discussed $\zeta$-function regularization [6].  In
all these different approaches to the self force there was significant
progress since the previous Capra meeting, although clearly much more
work is still needed.

A second, exciting direction which was emphasized for the first time
in the fourth Capra meeting, is the need for demonstrating the
connection between the different methods. This is important not just
in order to demonstrate the consistency and viability of the different
approaches, but also in order to compare their computational
effectivenesses, and perhaps even allow for a synergy of two or more
approaches. S.\ Detweiler proposed a (short) list of benchmark
problems, which he encouraged all the people who are working in this
field to consider, in order to confront and compare the different
approaches. Specifically, this list includes the problem of a scalar
charge in circular orbit around a Schwarzschild black hole, and the
calculation of gauge invariant quantities in the gravitational
analogue. It is hoped that much insight can be gained by such
comparisons. For the former benchmark problem one can already compare
the work by Nakano, Mino, and Sasaki [3] with earlier work
by Burko [7], and hopefully other researchers will
consider this problem too. A different way of comparing different
approaches is to compare the infinities which are
removed. Specifically, in the approach of Power-Expansion
regularization one computes the direct part of the self force.  In
Mode-Sum Regularization one typically computes the so-called
regularization function using local integrations of the Green's
function. As the two should agree, work is now in progress to show
just that.

Other interesting talks were given by A.\ Wiseman, who showed that the
self force on a static scalar charge in Schwarzschild spacetime is
exactly zero also when the scalar field is not minimally coupled
(despite earlier results by Zel'nikov and Frolov
[8], and by B.\ Whiting, who discussed how to
extend the Chrzanowski method to the time domain, and perhaps also
include sources. G.\ Schaefer described work with Damour and
Jaranowski on a post-Newtonian approach to radiation reaction, J.\
Levin discussed the fate of chaotic binaries, C.\ Glampedakis
described work with D.\ Kennefick on a `circularity theorem' for
spinning particles in Kerr spacetime, J.\ Pullin described a code in
the time domain to obtain radiation reaction waveforms, N.\ Andersson
described r-modes as a source of gravitational radiation, and C.\
Cutler described gravitational wave damping of neutron star
precession. S.\ Detweiler talked about gravitational self-force on a
particle in Schwarzschild spacetime, E.\ Poisson described work with
M.\ Pfenning on the self force in the weak-field limit
[9] , and B.\ Sathyaprakash discussed
resummation techniques for the binary black hole problem.

{\bf References:}

\noindent[1] Y.\ Mino, M.\ Sasaki, and T.\ Tanaka, Phys.\ 
Rev.\ D {\bf 55}, 3457 (1997)

\noindent[2]  T.\ C.\ Quinn and R.\ M.\ Wald, Phys.\
Rev.\ D {\bf 56}, 3381 (1997)

\noindent[3]  H.\ Nakano, Y.\ Mino and M.\ Sasaki, 
\htmladdnormallink{gr-qc/0104012}{http://xxx.lanl.gov/abs/gr-qc/0104012}

\noindent[4] L.\ Barack,
\htmladdnormallink{gr-qc/0105040}{http://xxx.lanl.gov/abs/gr-qc/0105040}

\noindent[5]  L.\ M.\ Burko and Y.\ T.\ Liu, Phys.\ Rev.\  D {\bf 64}, 024006
(2001)

\noindent[6] C.\ O.\ Lousto, Phys.\ Rev.\ Lett.\ {\bf 84}, 5251 (2000)

\noindent[7] L.\ M.\ Burko, Phys.\ Rev.\ Lett.\ {\bf 84}, 4529 (2000) 

\noindent[8] A.\ I.\ Zel'nikov and V.\ P.\ Frolov, 
Sov.\ Phys.\ JETP {\bf 55}, 191 (1982)

\noindent[9] M.\ J.\ Pfenning and E.\ Poisson,
\htmladdnormallink{gr-qc/0012057}{http://xxx.lanl.gov/abs/gr-qc/0012057}

\vfill\eject
\section*{\centerline {
Workshop on Numerical Relativity}\\ 
\centerline{Ngonyama, Krugersdorp Game Reserve}}
\addcontentsline{toc}{subsubsection}{\it  
Workshop on Numerical Relativity, by Michael Koppitz}
\begin{center}
Michael Koppitz, Albert Einstein Institute
\htmladdnormallink{koppitz@aei-potsdam.mpg.de}
{mailto:koppitz@aei-potsdam.mpg.de}
\end{center}

The Numerical Relativity 2001 workshop was organised as a
sister-conference to GR16, allowing many of the participants of that
conference to get involved in more in-depth discussions of the
numerical side of their work. As an added bonus, the workshop was held
at the \"Ngonyama Lion Lodge\" (http://www.afribush.co.za/) in a game
park near Johannesburg, with patrolling wildlife to ensure that
delegates did not stray too far from the conference venue.

After an introduction by Nigel Bishop, the stage was handed over to the
AEI/Golm group for the first session. Ed Seidel summarized work going on in
the EU-Network (http://www.eu-network.org) collaboration on numerical relativity,
and then focussed on work carried out at the AEI on colliding black holes.
Denis Pollney summarized gauge conditions which have proven crucial to
extending the life of evolutions both with and without excision. Carlos
Lousto and Manuela Campanelli closed the session off with a descriptionthe
Lazarus project, which combines full numerical studies of the plunge of two
black holes with the perturbative treatment starting in the linear regime.
They discussed the details of the idea, pointed out the inherent self-tests
this method provides how well it passes this tests, and showed recent
results obtained with this method, including the first full treatment of the
binary black hole system starting from the ISCO down to the ring down of the
final black hole. Later in the week, colliding black holes were also
demonstrated by Richard Matzner (Texas/Austin), who demonstrated his group's
grazing collision results.

Initial data methods were discussed over the afternoon. Peter Diener
(AEI/Golm) presented his results using an adaptive mesh technique
combined with a multi-grid solver to solve the initial data problem for
two black holes with the Kerr-Schild approach. Nina Jansen
(Tac/Copenhagen) used a similar solver to compare initial data sets in
terms of the behavior of the ADM mass as a function of separation.
Phillipe Grandclement (Observatoire de Paris/Meudon) presented a new
approach to the initial data problem, using multi-domain spectral
methods and a helical killing vector field to simplify the problem.
Micheal Koppitz (AEI/Golm) summarized the efforts underway at the AEI
to generate new sets of initial data for binary black hole systems,
emphasizing the strong need for comparison between sets.

The important role of constraints in the evolution equations was
emphasized in the interesting talks of Hisa-aki Shinkai (RIKEN/Japan)
and Oscar Reula (Cordoba).  Raymond Bursten, Anthony Lun, and
Elizabeth Stark (Monash University) described their use of the
electromagnetic parts of the Weyl tensor as well as the Bianchi
identities, to improve certain aspects of the conventional 3+1 evolution
system. Luis Lehner (British Columbia) also underlined the need for a
better understanding of boundary problems and non-principal terms of
the evolution equations.

Florian Siebel (MPI f\"ur Astrophysik/Germany) reported on fully
relativistic evolution of a neutron star using characteristic methods,
and Motoyuki Saijo (University of Illinois/Champaign) reported on
simulations of fate of the collapse of super-massive stars.  Marcelo
Salgado (UNAM/Mexico) presented a scalar-tensor model for neutron star
collapse.  Shin'ichirou Yoshida (SISSA/Trieste) described numerical
studies of rotation modes of differentially rotating neutron
stars. John Miller (SISSA/Trieste) gave an illuminating presentation
of the influence of shocks in domain significant to the r-mode
instability.

The advantages of the technique of Pad\'e approximants for
post-Newtonian calculation were described by Bala Iyer (Raman
Institute/Bangalore) Carsten Gundlach (Southampton) described the
methods used for his studies of critical collapse of perfect fluids.
Jose Martin-Garcia (Southampton) reported on a new formalism to
calculate nonspherical linear perturbations around a general spherical
background containing a perfect fluid. It is independent of the
equation of state and can therefore model physically interesting
problems.  David Hobill (University of Calgary) presented recent
studies of both sub-and super-critical Brill waves emphasising the
creation and evolution of trapped surfaces for super-critical initial
data.  Mihai Bondarescu (AEI/Golm) presented his work on embeddings of
2d surfaces (in particular, apparent horizons) in Minkowski space.

On the technical front, Garielle Allen and Thomas Radke (AEI/Golm)
gave a two hour introductory tutorial on the Cactus Computational
Toolkit, its underlying idea and benefits of using it even for small
scale computers and laptops. They explained how to obtain, compile
and run Cactus and presented some of its features, especially its
integration with visualization tools.  During the open discussion
session, adaptive mesh refinement (AMR) was identified as a urgent
requirement for bringing numerical relativity into the realm of real
physics problems, and it was good to see AMR work being carried out on
a number of fronts. In addition to the work of Diener and Jansen
already mentioned, Scott Hawley (AEI/Golm) presented his
implementation of Berger-Oliger type mesh refinements. Dae-Il Choi
(NASA Space Flight Center/Houston) showed AMR evolutions of strong
Brill waves demonstrating how the refined grid followed the wave very
well.

Alternative approaches to 3+1 were also discussed. Nigel Bishop (University
of South Africa/Pretoria) reported on results obtained by evolving a neutron
star orbiting around a Schwarzschild black hole using characteristic
techniques. Ray d'Inverno (Southampton) summarized recent work on
Cauchy-Characteristic matching techniques, in particular cosmic string
evolutions and progress on an axisymmetric code. Ruth Williams (Cambridge)
discussed discrete techniques involving space-times tessellated by polygons.
Carlos Sopuerta (University of Portsmouth) suggested a technique of using a
background metric to do the 3+1 split of Einstein's equations in situations
where the system under study is sufficiently known already. Osvaldo Moreschi
(Cordoba) described work on Robinson-Trautman space-times, for perturbations
specified at null infinity.  J\"org Frauendiener (Universit\"at
T\"ubingen/T\"ubingen) discussed boundary conditions, evolution schemes, and
technical problems arising when implementing the conformal field equations.

Sash Husa (AEI/Golm) reported on the current status of the
treatment of the conformal field equations and discussed some possible
future strategies. (Notable attendees to this latter session were the
conference site's resident hippos, who had until then refused to make an
appearance but listened attentively to the final day's talks from across the
pool.)

All in all it was an enjoyable week, not only for the physics that
were discussed, but also the fun location and Nigel Bishop and the
local organisers should be commended for putting it all together.

Denis Pollney was of great help in preparing this report.

\vfill\eject
\section*{\centerline {
Workshop on Canonical \& Quantum Gravity III}}
\addcontentsline{toc}{subsubsection}{\it  
Workshop on Canonical \& Quantum Gravity III by  J. Lewandowski and
J. Wi\'sniewski}
\begin{center}
 J Lewandowski (Warszaw), J. Wi\'sniewski (Penn State)\\
\htmladdnormallink{lewand@aei-potsdam.mpg.de}
{mailto:lewand@aei-potsdam.mpg.de}
\htmladdnormallink{jacek@gravity.phys.psu.edu}
{mailto:jacek@gravity.phys.psu.edu}
\end{center}

For a third time, a sizable portion of the gravity community
gathered in Warsaw to discuss recent advances.  Participation of
excellent physicists working on exciting topics, not limited just
to the field of canonical or quantum gravity, created a vibrant
atmosphere in the meeting. The workshop was sponsored by Banach
Center of Mathematics of Polish Academy of Sciences (PAN). It was
organized by Jerzy Lewandowski, Jacek Jezierski (Warsaw
University), Jerzy Kijowski (Center for Theoretical Physics PAN,
Warsaw) and Abhay Ashtekar (Penn State) who served as a scientific
advisor. Over 90 participants from 12 countries attended about 60
talks. Workshop was divided into two parts. First week was devoted
to problems in classical general relativity; its title was "Null
Structures and other Aspects of Classical Gravity". Second week
was devoted to problems of ``Quantum Gravity''. In between the two
parts was a one day celebration of Ted Newman's birthday, with
talks by Aichelburg, Ashtekar, Penrose, Stachel and Trautman as
well as interesting after dinner reminiscences in the evening, in
Palac Staszica. The organization of that day was directed by
Bialynicki-Birula, with the help  of  Demianski, Nurowski, Tafel
and Trautman.

By far the most extensive application of the null-cone structures
being the subject of the first part of the meeting is the Null
Surface Formulation (NSF) of the Einstein's theory started about
10 years ago by Newman and his collaborators. According to this
theory, the space-time is a secondary object defined as the set of
solutions of certain 3rd order ODE. The recent development
indicates relations with Cartan's theory of differential equations
(Newman, Nurowski, Kozameh) and applications to the gravitational
lensing (Fritelli, Tod). NSF is a relative of the Twistor program,
the advantage of the NSF being that it applies to real, not
necessarily analytic space-time of the Lorentz signature. However,
in one of his three talks Penrose reported on his recent attempt
to construct the twistor space for a generic curved space-time of
the $(+, -, -, - )$ signature! An exciting application of the
twistors that bridges the classical and the quantum theories is
the Bialynicki-Birula twistor Wigner-function introduced for the
participants of the workshop by its inventor. Twistor spaces
corresponding to anti-self-dual metrics in the $(+, +, -, - )$
signature with covariantly parallel spinor were characterized
before the audience by Dunajski.  A recent discovery of Damour,
Henneaux, Julia and Nicolai traces the roots of the BKL BKL
behavior near space-time singularities (in the dimensions
greater/equal 4) to the structure of the fundamental Weyl chamber
of some underlying hyperbolic Kac-Moody algebra. This intriguing
result and its consequences were presented in a comprehensive
lecture by Henneaux, one of the three talks on singularities
(Bizon, Aichelburg). Another major topic was the novel,
quasi-local generalization of the black hole theory, provided by
``isolated horizons'' (IHs). The mechanics and geometric
invariants including geometric conditions that distinguish the
Kerr horizon among all IHs were discussed (Beetle, Krishnan,
Lewandowski, Pawlowski). An interesting result shown by Racz was
his proof of the existence of a Killing vector in the case of the
bifurcate IHs. Related formulations of the mechanics of the null
shells and scri were explored by Chrusciel, Kijowski and Tafel. In
the area of the traditional black hole theory, Jacobson argued
that ``black hole entropy is not about black holes''. The recent
progress in understanding of the Penrose inequality was discussed
by Frauendiener. The ``canonical'' theme of the workshop was
underlined by the Beig's talk on the motion of the point particles
in general relativity and constraint equations. Other subject
covered were ``32 Double Coverings of $O(p,q)$ for $p,q>1$''
(Trautman),  ``The Hopf fibration -- five times in physics''
(Urbantke) and  ``Real Sources of Holomorphic Coulomb Fields''
(Kaiser).

The main topic discussed in the second, quantum, part of the
Workshop was ``quantum geometry''. Recent advances within this
approach in meeting that challenges of quantum gravity were
reviewed by Ashtekar in his lecture on the Newman day.  A focal
point of research in the canonical approach during last several
years has been the semi-classical sector of the theory.  Thiemann
and his collaborators (Winkler, Sahlmann) explored the idea of
construction of semi-classical states by gluing the coherent
states defined on SU(2). Another approach follows from
Varadarajan's embedding of the free Maxwell theory Fock space into
the U(1) analog of the polymer-like excitations Hilbert space of
the quantum geometry. A generalization to the SU(2) theory
described in the second of Ashtekar's talk provides a natural
candidate for the Fock flat space-time vacuum and a starting point
to bridge the background independent, non-perturbative approach
and perturbative results. A third way to extract a semi-classical
information from the non-perturbative sector was the subject of
Bojowald's talk. By a quantum symmetry reduction, and by
exploiting discreteness of volume in quantum geometry, he obtained
a substitute for the familiar Wheeler-DeWitt equation that
naturally resolves the big-bang singularity.  A second focal point
was provided by the lively discussions on ``spin foam models'',
which provide a path integral approach to the quantum gravity,
based again on quantum geometry.  The idea initiated by
Reisenberger and Rovelli and has drawn a great deal of attention
because of the recent finiteness results by Perez, Crane and
Rovelli which, roughly speaking, are analogous to the finiteness
claims of perturbative string theory. All principal researchers
(except Baez and Crane) in the area reported on the status of
their work. The promising idea of providing the space of the
spin-foams with the Hopf algebra structure was explained by
Markopoulo. The issue of observables in quantum geometry was
discussed by Pullin.  A third focal point to the workshop was
provided by simplicial Lorentzian gravity. Many of the frequently
asked questions were exhaustively answered by Loll, Ambjorn and
Jurkiewicz.

To provide a balance, there were several talks on the nearby
areas, particularly quantum groups (Woronowicz, Zapata,
Kowalski-Glikman), branes (Meisner, Pawelczyk, Louko), 2+1-gravity
(Bengtsson, Freidel, Wisniewski), the theta functions (Mourao), as
well as the talks on the status of the other issues of the quantum
theory, such as general covariance (Fredenhagen), gravitational
quantum state reduction (Penrose), gravitational collapse
(Hajicek), technical and conceptual issues in approaches based on
histories (Dasgupta, Kuchar), QCD on the lattice (Kijowski),
pre-canonical quantization (Kanatchikov). Especially instructive
was the lecture by Woronowicz on the representations of the
quantum Lorentz group. Those of us who apply the quantum groups in
everyday work, could ask the master about some subtleties and
other possible ways of q-deforming.

In summary, the atmosphere of the meeting was most stimulating due
to active participation of both, experienced as well as young
researchers. Interactions between different areas of research in
mathematical/quantum gravity and at the same time avoiding the
overload of big conferences seemed to be an advantage. Hopefully,
the Warsaw workshops CQG have already become a tradition and
future ones will again bring excellent researchers and lecturers.

\end{document}